\documentclass[%
 aip,
 amsmath,amssymb,
 reprint,%
]{revtex4-1}

\usepackage{graphicx}
\usepackage{dcolumn}
\usepackage{bm}
\usepackage[dvipsnames]{xcolor}
\usepackage[utf8]{inputenc}
\usepackage[T1]{fontenc}
\usepackage{mathptmx}
\usepackage{etoolbox}

\makeatletter
\def\@email#1#2{%
 \endgroup
 \patchcmd{\titleblock@produce}
  {\frontmatter@RRAPformat}
  {\frontmatter@RRAPformat{\produce@RRAP{*#1\href{mailto:#2}{#2}}}\frontmatter@RRAPformat}
  {}{}
}%
\makeatother

\begin{document}

\preprint{AIP/123-QED}

\title[In-situ uniaxial pressure cell for X-ray and neutron scattering experiments]{In-situ uniaxial pressure cell for X-ray and neutron scattering experiments}
\author{G.~Simutis}
\email{marc.janoschek@psi.ch}
\affiliation{Laboratory for Neutron and Muon Instrumentation,
Paul Scherrer Institut, CH-5232 Villigen PSI, Switzerland}
\affiliation{Department of Physics, Chalmers University of Technology, SE-41296 G\"oteborg, Sweden}

\author{A.~Bollhalder}
\email{alex.bollhalder@psi.ch}
\affiliation{Laboratory for Neutron and Muon Instrumentation,
Paul Scherrer Institut, CH-5232 Villigen PSI, Switzerland}

\author{M.~Zolliker}
\affiliation{Laboratory for Neutron and Muon Instrumentation,
Paul Scherrer Institut, CH-5232 Villigen PSI, Switzerland}

\author{J.~K\"uspert}
\affiliation{Physik-Institut, Universit\"{a}t Z\"{u}rich, Winterthurerstrasse 190, CH-8057 Z\"{u}rich, Switzerland}

\author{Q.~Wang}
\affiliation{Physik-Institut, Universit\"{a}t Z\"{u}rich, Winterthurerstrasse 190, CH-8057 Z\"{u}rich, Switzerland}

\author{D. Das}
\affiliation{Laboratory for Muon Spin Spectroscopy, Paul Scherrer Institute, Villigen PSI, Switzerland}

\author{F. Van Leeuwen}
\affiliation{Laboratory for Neutron and Muon Instrumentation, Paul Scherrer Institut, CH-5232 Villigen PSI, Switzerland}
\affiliation{Structure and Mechanics of Advanced Materials, Paul Scherrer Institut, CH-5232 Villigen PSI, Switzerland}

\author{O. Ivashko}
\affiliation{Deutsches Elektronen-Synchrotron DESY, Notkestraße 85, 22607 Hamburg, Germany}

\author{O. Gutowski}
\affiliation{Deutsches Elektronen-Synchrotron DESY, Notkestraße 85, 22607 Hamburg, Germany}

\author{J. Philippe}
\affiliation{Laboratory for Neutron and Muon Instrumentation,
Paul Scherrer Institut, CH-5232 Villigen PSI, Switzerland}
\affiliation{Physik-Institut, Universit\"{a}t Z\"{u}rich, Winterthurerstrasse 190, CH-8057 Z\"{u}rich, Switzerland}

\author{T. Kracht}
\affiliation{Deutsches Elektronen-Synchrotron DESY, Notkestraße 85, 22607 Hamburg, Germany}

\author{P. Glaevecke}
\affiliation{Deutsches Elektronen-Synchrotron DESY, Notkestraße 85, 22607 Hamburg, Germany}

\author{T. Adachi}
\affiliation{Department of Engineering and Applied Sciences, Sophia University, Chiyoda, Tokyo 102-8554, Japan}

\author{M. Von Zimmermann}
\affiliation{Deutsches Elektronen-Synchrotron DESY, Notkestraße 85, 22607 Hamburg, Germany}

\author{S. Van Petegem}
\affiliation{Structure and Mechanics of Advanced Materials, Paul Scherrer Institut, CH-5232 Villigen PSI, Switzerland}

\author{H. Luetkens}
\affiliation{Laboratory for Muon Spin Spectroscopy, Paul Scherrer Institute, Villigen PSI, Switzerland}

\author{Z. Guguchia}
\affiliation{Laboratory for Muon Spin Spectroscopy, Paul Scherrer Institute, Villigen PSI, Switzerland}

\author{J.~Chang}
\affiliation{Physik-Institut, Universit\"{a}t Z\"{u}rich, Winterthurerstrasse 190, CH-8057 Z\"{u}rich, Switzerland}

\author{Y.~Sassa}
\affiliation{Department of Physics, Chalmers University of Technology, SE-41296 G\"oteborg, Sweden}

\author{M. Bartkowiak}
\affiliation{Laboratory for Neutron and Muon Instrumentation,
Paul Scherrer Institut, CH-5232 Villigen PSI, Switzerland}

\author{M. Janoschek}
\email{gediminas.simutis@psi.ch}
\affiliation{Laboratory for Neutron and Muon Instrumentation,
Paul Scherrer Institut, CH-5232 Villigen PSI, Switzerland}
\affiliation{Physik-Institut, Universit\"{a}t Z\"{u}rich, Winterthurerstrasse 190, CH-8057 Z\"{u}rich, Switzerland}

\date{\today}

\begin{abstract}
We present an in-situ uniaxial pressure device optimized for small angle X-ray and neutron scattering experiments at low-temperatures and high magnetic fields. A stepper motor generates force, which is transmitted to the sample via a rod with integrated transducer that continuously monitors the force. The device has been designed to generate forces up to 200 N in both compressive and tensile configurations and a feedback control allows operating the system in a continuous-pressure mode as the temperature is changed. The uniaxial pressure device can be used for various instruments and multiple cryostats through simple and exchangeable adapters. It is compatible with multiple sample holders, which can be easily changed depending on the sample properties and the desired experiment and allow rapid sample changes.
\end{abstract}

\maketitle

\section{\label{sec:Intro}Introduction}

In modern condensed matter physics, understanding and control of collective quantum behavior is at the forefront of fundamental research. At the same time, it is the driver for future technologies with potential applications of quantum matter ranging from quantum computation and cryptography to energy harvesting and dissipationless electricity transfer in superconductors.\cite{Tokura2017,Giustino2020,Cava2021, Marsden2021}

Much of the interest in quantum matter arises from the ability to efficiently tune and switch their properties using external perturbations. This 'tunability' is a result of quantum phases typically emerging due to delicate interplay of several atomic-scale interactions .\cite{Vojta2003,Sachdev2011} Unconventional superconductivity is perhaps the best-known example of a macroscopic quantum phase. Notably, extensive experimental studies carried out over the last few decades have highlighted that many quantum materials may be tuned towards a zero-temperature magnetic instability using external control parameters such as chemical substitution, hydrostatic pressure or magnetic field, where the associated quantum fluctuations are generally believed to mediate superconductivity.\cite{Scalapino2012}

Over the last decade, and in addition to these well-established tuning parameters, uniaxial pressure (or strain) has been established as a crucial tool, which aided in revealing and understanding a series of remarkable new quantum states.  For example, it was exploited to unveil the complex superconducting pairing symmetry of Sr$_2$RuO$_4$.\cite{Hicks2014,Steppke2017,Sunko2019}. Most recently, uniaxial pressure has enabled studying the nature of the charge and spin density waves as well as their interplay with superconductivity in cuprates. In YBa$_2$Cu$_3$O$_{7-x}$, uniaxial pressure was found to induce the zero-magnetic-field three-dimensional charge ordering,\cite{Kim2018,Kim2021} while in {La}$_{2-x}${Sr}$_{x}${CuO}$_4$ it enables repopulation of the stripe order domains.\cite{Choi2022,Wang2022,Simutis2022} In {La}$_{2-x}${Ba}$_{x}${CuO}$_4$ an extremely low uniaxial stress induces a threefold increase in the onset of three dimensional superconductivity. The 3D superconducting coherence was shown to be anti-correlated with large-volume-fraction spin-stripe order and the relative prominence of the two phases can be regulated by the uniaxial pressure.\cite{Guguchia2020,Kamminga2022} In a broader view of quantum materials, strain has also been suggested to stabilize exotic phases ranging from the enigmatic ‘hidden order’\cite{Bourdarot2011} to quantum spin liquids \cite{Kaib2021} to skyrmion lattices,\cite{Nii2015,Chacon2015}, which are thought to be relevant for novel spintronics and memory applications.\cite{Fert2013}

Recently there has been tremendous progress in lab-based uniaxial devices based on piezoelectric stacks. In particular devices based on the developments described in Ref.~\onlinecite{Hicks2014b} have been commercialized and used in laboratories worldwide. Nevertheless, to further access the subtle details of the relevant quantum phases and to disentangle the complex interplay of spin, charge and lattice degrees of freedom, experiments at X-ray synchrotron, muon and neutron sources are often required. Various uniaxial pressure cells were optimized for different scattering geometries and successfully used, with designs ranging from differential thermal expansion\cite{Boyle2021} to one-screw compression \cite{Choi2022} to an adaptation of anvil-type cells\cite{Edberg2020} without any transmission medium. A major drawback of these cells is that they only allow applying pressure ex-situ, which is time consuming and inefficient.

Therefore, given the strong demand for the measurement time at large scale facilities, complex sample environments, such as uniaxial pressure devices must be highly efficient and hence in-situ tuning is the necessary approach. An additional, but relevant, advantage of in-situ system is that it allows the isothermal tuning of uniaxial pressure. Moreover, the samples are often larger than in lab-based measurements, which can further push the complexity of the design. Two approaches have been so far employed. To adapt the piezoelectric system to allow for bigger displacements, a piezoelectric stack device with multiple elements was built\cite{Ghosh2020} and successfully used for muon spin rotation/relaxation experiments.\cite{Guguchia2020, Grinenko2021} Alternatively, copper bellows can be filled with pressurized helium gas to generate the force needed for manipulating the materials.\cite{Fobes2017,Degiorgi2022} While very successful, both of these approaches require extensive complexity and nontrivial operation.

The device presented here is a new solution, based on simple operating principles, yet providing accurate application of pressure, the ability to apply force in-situ at temperatures ranging from room down to cryogenic temperatures. The device is also sufficiently compact to even fit into a cryomagnet, allowing for simultaneous tuning of temperature, magnetic field, and uniaxial pressure. Furthermore, the option to quickly exchange the premounted samples and the availability of multiple modular sample holders optimize the precious measurement time at large scale facilities. The chapters below outline the operating principle and the details of the device followed by test measurements showing the first results of the operational system.

\section{Technical implementation of the device}
\subsection{\label{sec:Force}Force generation and transmission}

\begin{figure}
\includegraphics[width={\columnwidth}]{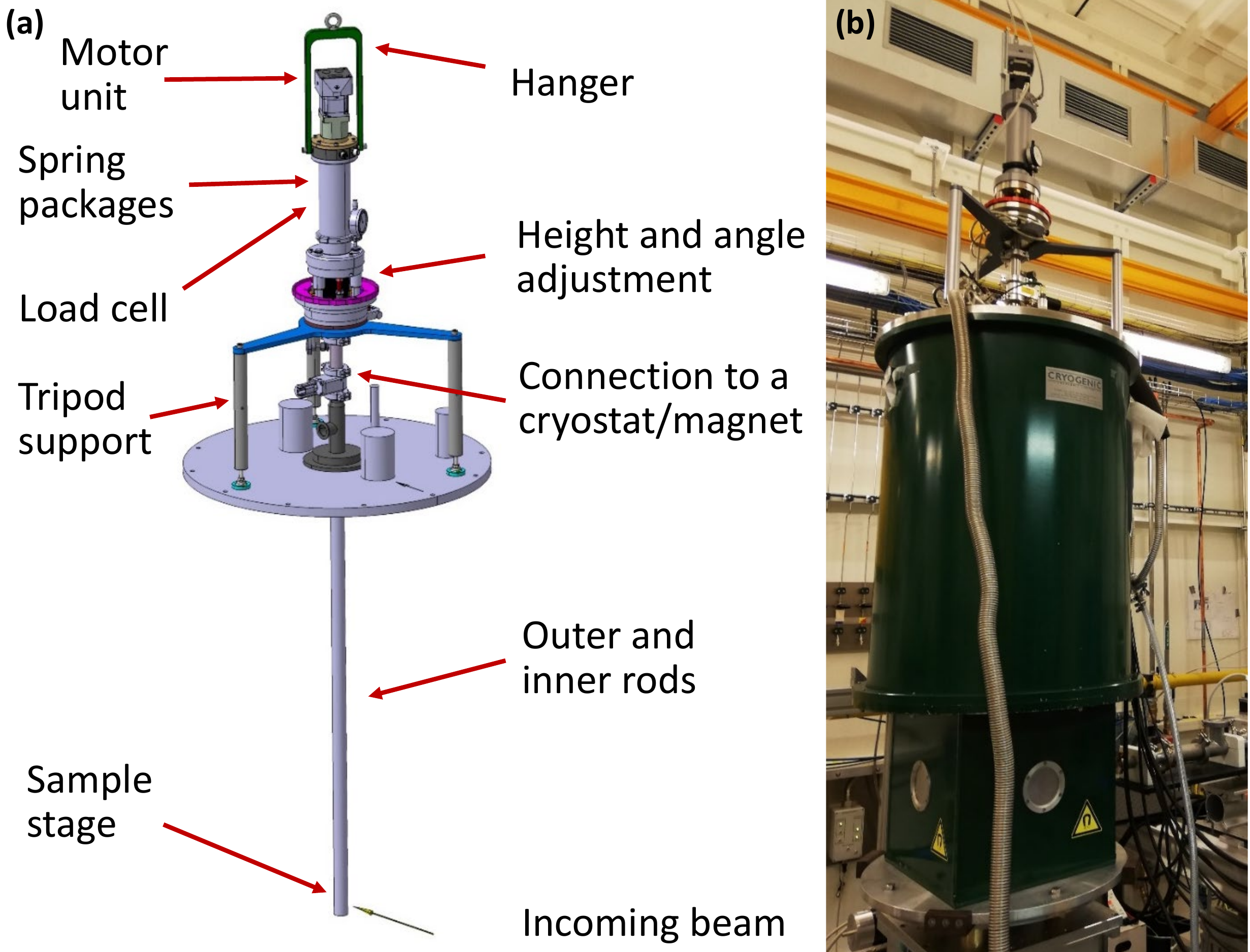}
\caption{\label{fig:Overview} The new uniaxial pressure device. (a) indicates the the main parts of the force generation and transmission elements of the pressure cell. (b) shows the pressure device positioned in a CRYOGENIC horizonthal-10T cryomagnet on the P21.1 beamline in PETRA-III synchrotron at DESY.}
\end{figure}

A general overview of the apparatus is provided in Fig.~\ref{fig:Overview}. In order to achieve precise control, the force in our device is generated using a motor linear actuator (MLA) with microstepping from Trinamic company (PD57-2-1161). The MLA allows for smallest available rotational step size of 0.00703 degrees, corresponding to a travelling distance of 0.0195 $\mu$m. While this is already a fine step, for our purposes of a maximal strain of about 1\% of a typical sample of $\thicksim$ 2 mm, the full movement of 0.02 mm corresponds to only $\thicksim$ 1000 steps.

To increase the resolution, additional spring disc packages are installed in series. They start deforming at low forces and consequently increase the effective linear movement. Moreover, this ensures that the force is applied gently after the engagement and preserves potentially fragile samples.

In our implementation we use four standard EN 16983 disc springs in series, each of which plastically deforms by 0.176 mm upon application of 200 N force, which is the maximum anticipated value for the experiments. This, combined with extra deformation of the load cell ($\thicksim$0.07 mm/200 N) increases the effective travelling distance up to $\thicksim$ 0.8 mm corresponding to $\thicksim$ 40000 microsteps, hence vastly improving the control of the precise load on the samples.

\begin{figure}
\includegraphics[width={\columnwidth}]{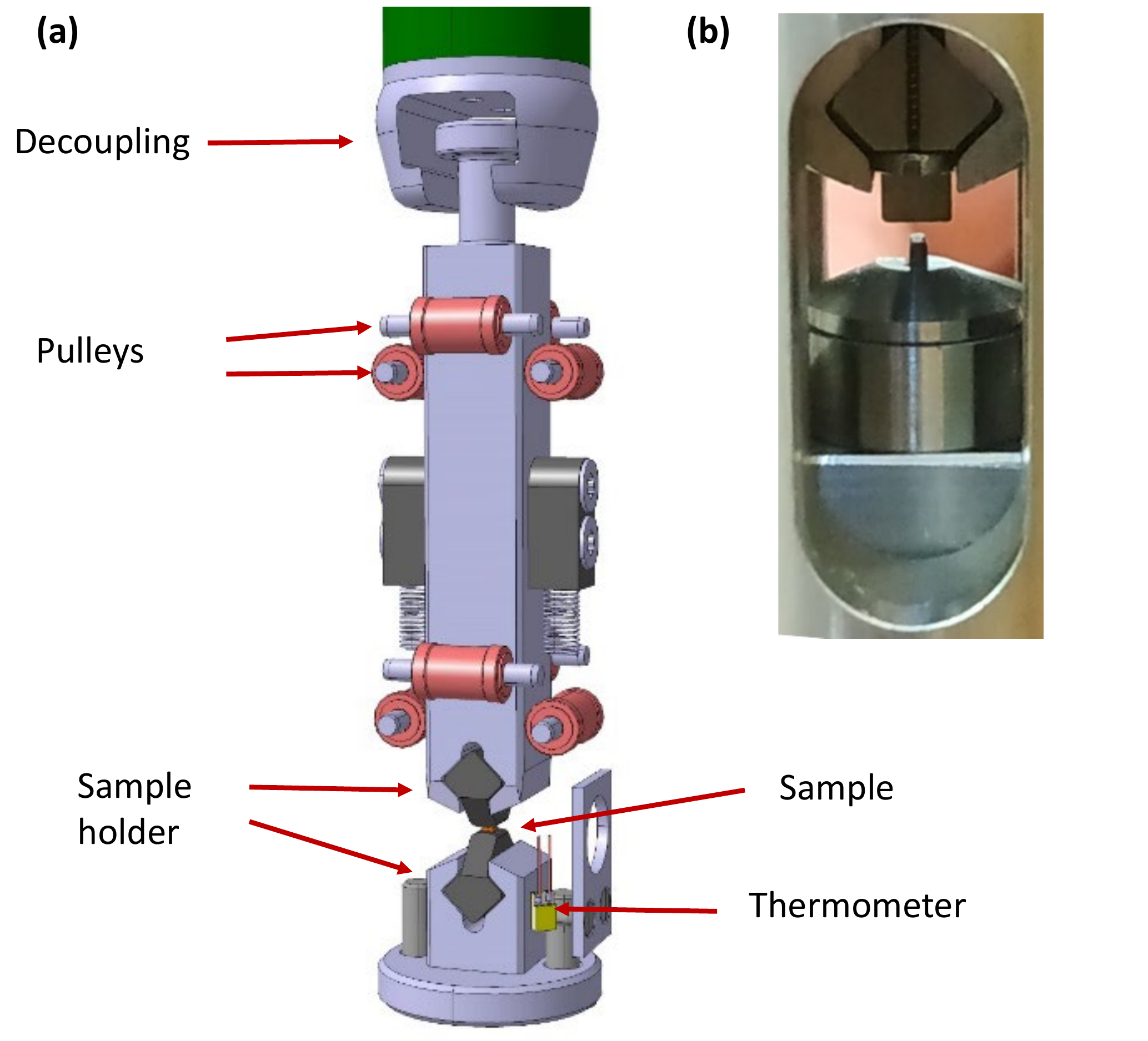}
\caption{\label{fig:Sample_stage} The sample stage of the apparatus (a) showing the force transmission parts, thermometer and sample positioned in the multi-directional holder. Panel (b) shows a La$_{2-x}$Ba$_x$CuO$_4$ sample installed for measurements at P21.1 beamline, using a push-only sample holder.}
\end{figure}

The 1.5 m long stainless steel (type 1.4404) push rod is connected to the sample stage (Fig.~\ref{fig:Sample_stage}), where the force is transferred to the sample holders. In order to ensure a straight and efficient transmission of force through the sample stage, it is guided by a set of pulleys. The rod and the sample stage can be mechanically decoupled - there is a free movement of the rod of 3 mm, with respect to the sample stage. This decoupling mechanism protects from any buildup of strain upon temperature variation and enables true zero pressure reference measurements. In order to apply the tensile (compressive) strain, the rod has to be retracted (extended) until the engagement. The moment of engagement can be traced via the load cell (see also below).

The sample stage terminates with the slots for sample holders as seen in Fig.~\ref{fig:Sample_stage}. The square outside shape of the sample holders ensures a proper alignment for pushing and pulling operations, while transmitting the applied force efficiently. Apart from the constraint of the square fixation shape, the sample stage can accept a variety of sample holders, some of which are presented in Section \ref{sec:Holders} and is readily adaptable for future implementations.

\subsection{\label{sec:Control}Feedback and Control}

The applied force is measured using a load cell (see Fig. 1) from Transducer Techniques (MLP 50) which allows measuring loads of up to +-220 N. In addition to the force measurement, the apparatus is measuring the temperature in the vicinity of the sample (see Fig. 2) using a Cernox thin film resistance cryogenic temperature sensor (CX-SD-PACKAGE), which is read in via LakeShore-340 controller.

The software control is realised with the Frappy package \cite{Kiefer2019} on an embedded server board (PC engines apu4d4). It enables processing the measured force and temperature values as well as sending commands to the motor to initiate or stop movement. For situations, when a constant-force measurement is required, a feedback loop can be enabled to compare the read in force with the desired value and apply corrective action using the motor. This mode is essential when large temperature variation takes place during the measurement.

\subsection{\label{sec:Holders}Sample Holders}

\begin{figure}
\includegraphics[width={\columnwidth}]{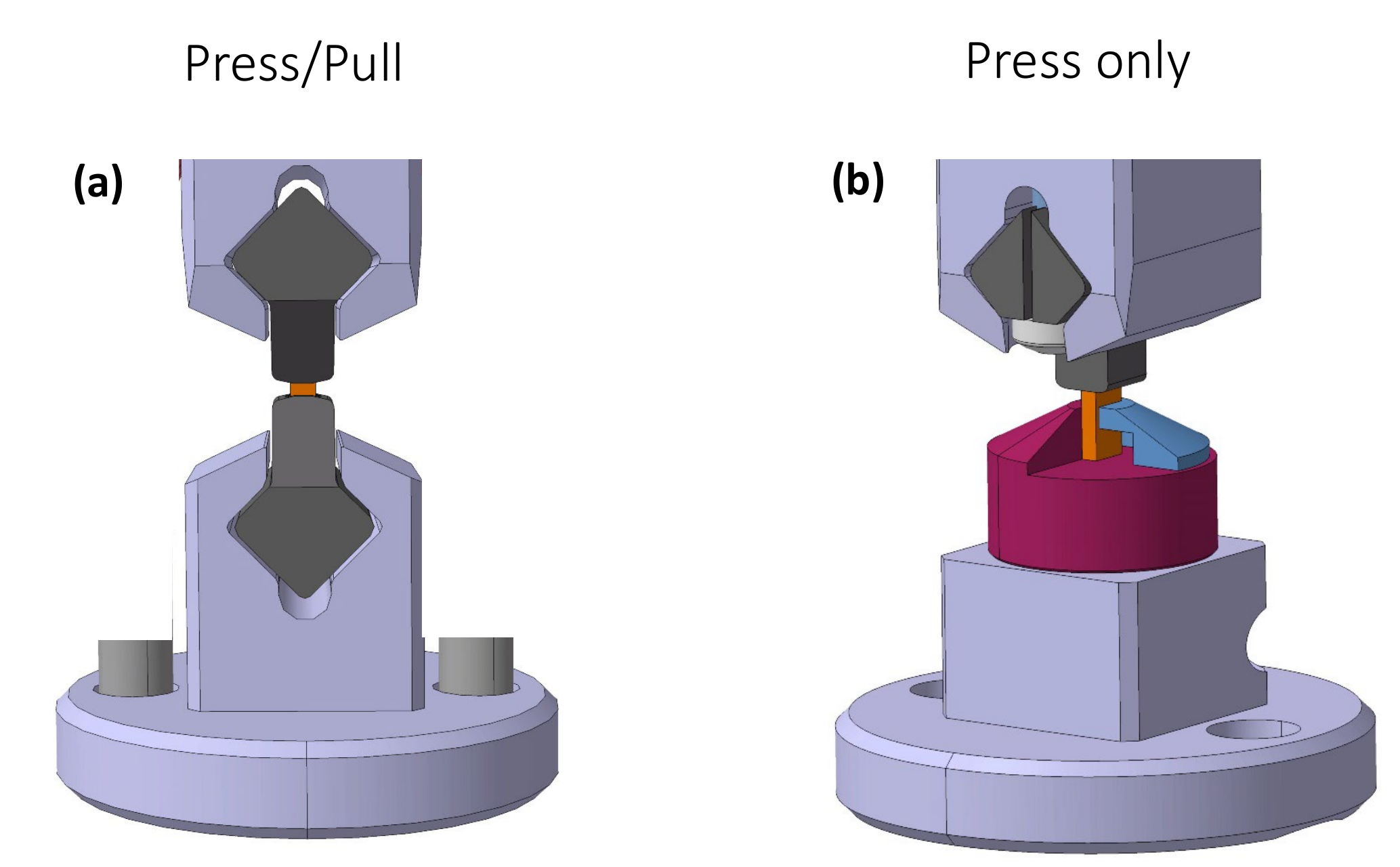}
\caption{\label{fig:Sample_holders} Different sample holders that have been used with the uniaxial pressure device. (a) shows the sample holder with both push and pull capacity, whereas (b) is optimized for samples, where pushing only is needed.}
\end{figure}

The present apparatus is able to apply compressive and tensile pressure, and consequently, the sample holders were designed to support both push and pull measurements. Moreover, the holders can be replaced quickly so that previously prepared and glued samples can be efficiently exchanged during the beamtime when measurement time is precious. The types of the holders can also be quickly changed by removing the bottom part of the sample stage.

Two sample holders, which were already successfully tested and deployed in experiments, are presented in Fig.~\ref{fig:Sample_holders}. For the holder compatible with both pushing and pulling ( Fig.~\ref{fig:Sample_holders}a), the sample is glued into the inside tubes of the holders using epoxy. The top and bottom sample holders are then simultaneously slid into the square slots, which maintains the whole system straight.

For samples, where only pushing action is needed, a different holder shown in Fig.~\ref{fig:Sample_holders}b can be used. The top part of the holder is replaced by a pusher, which is stabilized by screwing it into the sample stage. The bottom part has a versatile base, upon which various heads can be positioned, depending on the size of the sample and its properties. In the presented case, the sample (orange) is glued using epoxy to the bottom part of the sample. The bottom part of the sample can be rotated in order to access different parts of the reciprocal space, in case of limited scattering angle of the sample environment.

\section{\label{sec:SampEnv}Compatibility with instrumentation}
The uniaxial device presented here can be used in various scenarios, but is particularly suited for  experiments in a quasi-transmission geometry featuring small scattering angles, such as high energy (100 keV) x-ray diffraction and small angle neutron scattering (SANS) measurements. It is attached to the cryostat or magnet using a standard KF25 flange or using additional adapters when needed. While for small tilting angles this provides enough support, upon further tilting, or for situations with long adapters, an additional tripod was designed to prevent torque on the device (Figure 1).

In order to make sure that the sample position could be modified inside the cryostat for centering and signal optimization purposes, a displacement unit is installed that can translate the UPD along the vertical direction with a range of 18 mm.

Moreover, there is an additional radial adjustment system to align the oriented sample with the available windows in the magnets and cryostats.

Control-wise, the instrument has been designed to work using the SEA client-server software for controlling and monitoring sample environment devices at the Swiss Spallation Neutron Source SINQ. It has been further incorporated into the beamline control elements at PETRA-III synchrotron using PyTango device control framework.

\section{\label{sec:Experiment} Experimental tests}

\begin{figure}
\includegraphics[width={\columnwidth}]{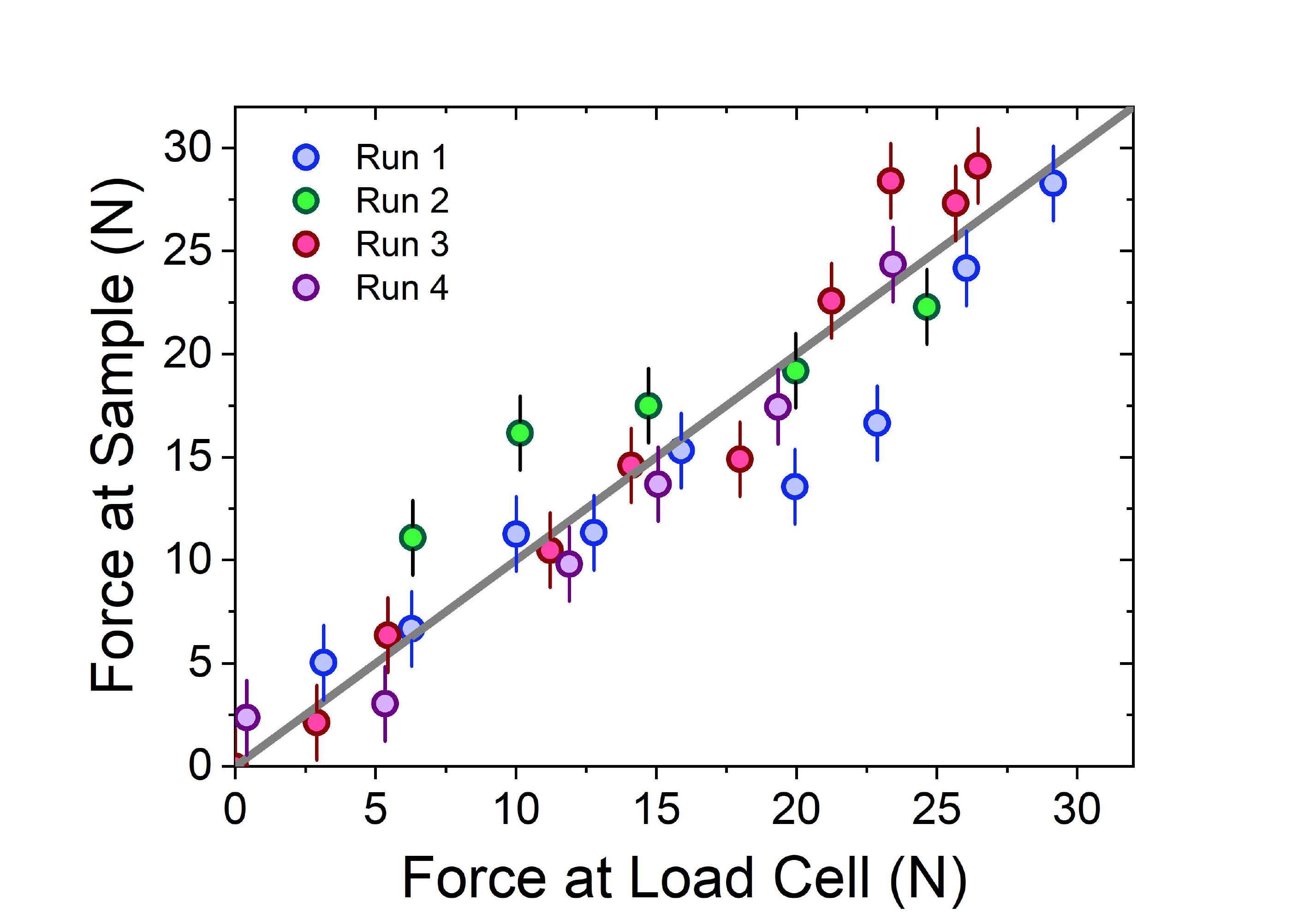}

\caption{\label{fig:MTM_Comparison} The measurement of the efficiency of the force transmission. The abscissa shows the force values as read in from the transducer, whereas the ordinate corresponds to the extracted force from the optical measurement of the length of a previously calibrated steel dogbone. Runs 1 and 3 are upon loading, Runs 2 and 4 are upon releasing the force.}
\end{figure}

In order to test the operation of the device and to check the efficiency of the force transfer through the long distance of the apparatus, we have first measured the stress-strain curves of a steel mini dogbone (t = 0.5 mm, d = 1.7 mm) using the setup described here and comparing it with the measurement of the exactly the same dogbone using a Micro Tensile Machine \cite{Swygenhoven2006}, where the load cell is right next to the sample, which in turn is right next to the stepper motor. This allowed us to compare the readings of the load cells of the two systems at identical strain values. As seen in the Fig.~\ref{fig:MTM_Comparison}, the agreement is very good, verifying the efficient force transfer across the device.

\begin{figure}
\includegraphics[width={\columnwidth}]{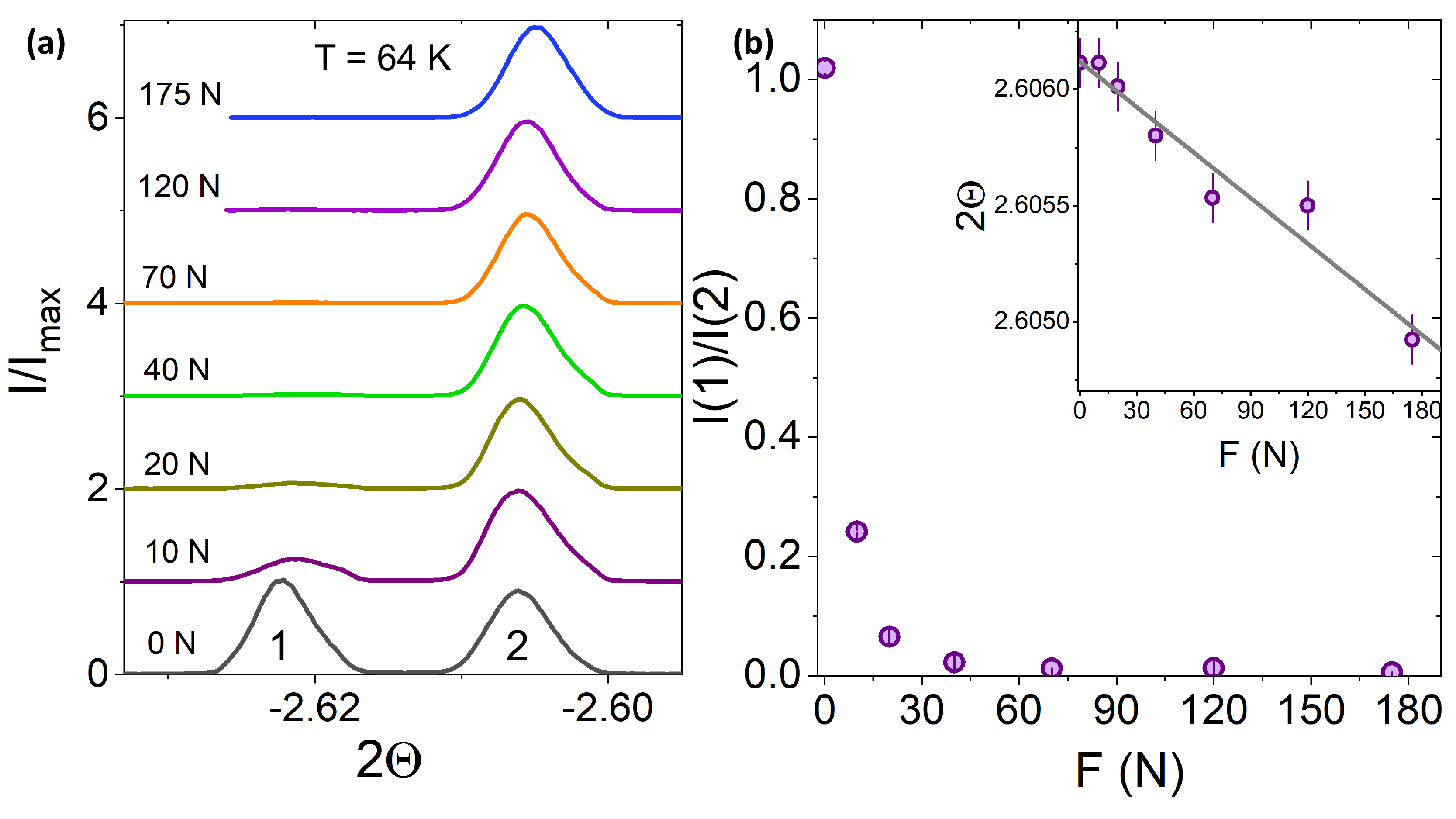}
\caption{\label{fig:LBCO_data} The effect of uniaxial pressure on the  structure of {La}$_{1.885}${Ba}$_{0.115}${CuO}$_4$. (a) shows the diffraction data on the structural peaks of {La}$_{1.885}${Ba}$_{0.115}${CuO}$_4$ at different applied forces. The peaks labelled "1" and "2" correspond to different structural domains in the orthorhombic phase. The intensity of one of the peaks is suppressed due to detwinning, with the intensity ratio of the two peaks shown in (b). The scattering angle of the surviving peak is continuously decreased, as expected for this geometry, and is plotted in the inset.}
\end{figure}

Having satisfied the requirement of efficient force transfer, we then checked whether the crystals are continuously deformed up to high forces while on the P21.1 beamline at PETRA-III. We have investigated a pertinent case of the {La}$_{2-x}${Ba}$_{x}${CuO}$_4$ system with x = 0.115. The full study of the properties under strain will be reported elsewhere \cite{Guguchia2022}, here we present the demonstration of the uniaxial pressure effects in the orthorhombic phase at 64 K. The crystal was positioned with tetragonal [0,0,1] and [1,1,0] directions spanning the scattering plane and the pressure applied perpendicular to the plane.    Fig.~\ref{fig:LBCO_data} shows how the structural peaks arising from the (1,1,0) tetragonal peak respond to the application of the uniaxial pressure.

Since the material is orthorhombic at these temperatures, two domains (corresponding to the (2,0,0) and (0,2,0) peaks in the orthorhombic notation) can be expected as seen in Fig.~\ref{fig:LBCO_data}a with no applied force. Uniaxial pressure has long been used to structurally detwin orthorhombic crystals and it is also observed in our case, with one of the peaks losing the intensity very quickly, also plotted in Fig.~\ref{fig:LBCO_data}b. In addition, we track the scattering angle of the surviving peak as we increase the applied force (inset of Fig.~\ref{fig:LBCO_data}b). As expected, when measuring the peak perpendicular to the applied force (i.e. in the expanding direction), we observe the reduction of the scattering angle.

Our measurements demonstrate that the force can be continuously applied with high precision and the structure of the crystals is efficiently modified, enabling tuning the properties of the material under investigation, in this case the {La}$_{2-x}${Ba}$_{x}${CuO}$_4$ system with x = 0.115.

\section{\label{sec:Summary} Summary}

In summary, we have designed and commissioned a new in-situ uniaxial pressure device for X-ray and neutron scattering experiments, specifically tailored for the investigation of quantum matter at low temperatures and high magnetic fields. In particular, the design of the sample holder is highly flexible and can be adapted to support various sample requirements. The control system designed for the system can be easily integrated in the control stack of large-scale research facilities. Taken together, the device can be easily used for a wide range of experimental situations.

\begin{acknowledgments}
The project has received funding from the European Union’s Horizon 2020 research and innovation program under the Marie Skłodowska-Curie grant agreement No 884104 (PSI-FELLOW-III-3i). G.S. acknowledges funding from the Chalmers X-Ray and Neutron Initiatives (CHANS). Y.S. acknowledges funding from the Swedish Research Council (VR) through a Starting Grant (Dnr.2017-05078) and from the Area of Advance - Material Sciences from Chalmers University of Technology. T.A. was supported by JSPS KAKENHI through Grant No. 19H01841. J.~K., Q.~W., and J.~C. acknowledge support from the Swiss National Science Foundation (200021$\_$188564). J.K. is further supported by the PhD fellowship from the German Academic Scholarship Foundation. We acknowledge DESY (Hamburg, Germany), a member of the Helmholtz Association HGF, for the provision of experimental facilities. Parts of this research were carried out at beamline P21.1 of PETRA II synchrothron. Beamtime was allocated for long-term project proposal II-20190002 EC and a regular proposal I-20210712 EC.

\end{acknowledgments}

\section*{AUTHOR DECLARATIONS}
\subsection*{Conflict of Interest}

The authors have no conflicts to disclose.

\section*{Data Availability Statement}

The data that support the findings of this study are available from the corresponding author upon reasonable request.

\section*{References}

\nocite{*}
\bibliography{straincell}

\end{document}